# Neuromorphic Robust Estimation of Nonlinear Dynamical Systems Applied to Satellite Rendezvous

Reza Ahmadvand, Sarah Safura Sharif, Yaser Mike Banad[*]
*School of Electrical and Computer Engineering, University of Oklahoma, Oklahoma, 73019, U.S.A.*
Emails: iamrezaahmadvand1@ou.edu, s.sh.@ou.edu, bana@ou.edu
*Corresponding author

**Abstract**

State estimation of nonlinear dynamical systems has long aimed to balance accuracy, computational efficiency, robustness, and reliability. The rapid evolution of various industries has amplified the demand for estimation frameworks that satisfy all these factors. This study introduces a neuromorphic approach for robust filtering of nonlinear dynamical systems: SNN-EMSIF (spiking neural network-extended modified sliding innovation filter). SNN-EMSIF combines the computational efficiency and scalability of SNNs with the robustness of EMSIF, an estimation framework designed for nonlinear systems with zero-mean Gaussian noise. Notably, the weight matrices are designed according to the system model, eliminating the need for a learning process. The framework's efficacy is evaluated through comprehensive Monte Carlo simulations, comparing SNN-EMSIF with EKF and EMSIF. Additionally, it is compared with SNN-EKF in the presence of modeling uncertainties and neuron loss, using RMSEs as a metric. The results demonstrate the superior accuracy and robustness of SNN-EMSIF. Further analysis of runtimes and spiking patterns reveals an impressive reduction of 85% in emitted spikes compared to possible spikes, highlighting the computational efficiency of SNN-EMSIF. This framework offers a promising solution for robust estimation in nonlinear dynamical systems, opening new avenues for efficient and reliable estimation in various industries that can benefit from neuromorphic computing.

**Keywords**: Neuromorphic computing, Spiking neural network, Sliding innovation filter, Kalman filter, Robust estimation

## 1. Introduction

Computational efficiency, robustness, and reliability have long been dominant factors for implementing estimation frameworks in the front-end of navigation systems. Estimation often means extracting meaningful data from the noisy measured signals provided by various types of sensors to the aims such as denoising (Obidin & Serebrovski, 2014), and sometimes accurate estimation of unmeasured states/parameters of a system (Revach, et al., 2022) which is the concern of the present study. Though, the popular estimation strategy called Kalman filter (KF) (Kalman, 1960) which has a wide range of applications in different scientific areas such as radar signal processing (Roth, et al., 2017; Jung, et al., 2020), spacecraft state estimation (Lefferts, et al., 1982; Filipe, et al., 2015), vision-based navigation (Kawamura, et al., 2022; Kawamura, et al., 2023), fault-identification in autonomous systems (Fekrmandi, et al., 2023), and even denoising of the signals obtained by fiber-optic gyroscopes (Narasimhappa, et al., 2012), it has been proposed for the optimal estimation of well-defined deterministic systems with linear dynamical models. Thus, through the analytical linearization leveraging the computation of Jacobian matrices of the dynamic and measurement model, at the cost of accuracy, the extended KF (EKF) has been introduced to estimate nonlinear systems (Crassidis & Junkins, 2011). However, from the robustness perspective, since the gain formulation of the KF family is developed based on minimizing the variance of the state estimation error, they struggle when encountering any form of uncertainty in the implemented model. This can lead to a degradation in accuracy and, in some cases, cause the estimation process to become unstable and diverge. Consequently, different variations of the KF with enhanced robustness, such as two-stage EKF (Xiao, et al., 2015) and inspired by the strong tracking filter (STF), an adaptive fading KF (AFKF) have been proposed by Xiao et al. (Xia, et al., 1994). Notably, the AFKF benefits from the fading memory concept to improve the robustness of the filter against the modeling errors in the linear systems by increasing the importance of the recent observations compared to the old ones. Further, to encounter the uncertainty in the nonlinear systems, the combinations of the STF with the unscented KF (UKF) and cubature KF (CKF) that are different versions of Gaussian approximation filters that leverage sigma-points instead of analytical linearization have been proposed (Gourabi, et al., 2020; Liu & Wu, 2017; Zhang, et al., 2018). In addition to the aforementioned strategies, another robust minimax estimator called H∞ filter, which minimizes the the state estimation error in the worst-case and can be considered as a robust variant of KF, has been introduced (Banavar, 1992; Zhao & Mili, 2018). However, the H∞ filter suffers from a complicated fundamental theory and a great performance sensitivity to tuning parameters (Simon, 2006). Therefore, on the contrary to the KF's minimum mean square error estimation algorithm to relax the drawbacks of KF in the presence of uncertainties, based on the sliding mode theories and exploitation of the sliding mode observers, as the first sliding estimator which implements the *prediction-correction* algorithm the variable structure filter (VSF) has been proposed for linear systems (Habibi & Burton, 2003). In 2007, the extended version of the VSF called the smooth variable structure filter (SVSF) was proposed, and its superior performance compared to the popular filters of the KFs family and particle filter (PF) in the presence of the modeling uncertainty and initial condition error has been demonstrated (Habibi, 2007) (Gadsden, et al., 2009; Wang, et al., 2008; Al-Shabi, et al., 2013). Next, through to the definition of a time-varying boundary layer for SVSF (Gadsden, 2011) the combined forms of EK-SVSF, UK-SVSF and CK-SVSF have presented to the heritage the robustness of the SVSF simultaneously with the optimality of KF. Then, in comparative studies, the superior performance of CK-SVSF has been established (Gadsden, 2011; Gadsden, et al., 2014; Gadsden, et al., 2014). In 2020, the sliding innovation filter (SIF), considered the next generation of SVSF, was introduced for uncertain linear systems with Gaussian noises (Gadsden & Al-Shabi, 2020). Subsequently, implementing Jacobians, the SIF formulation has been extended to nonlinear systems, referred to as the extended SIF (ESIF).

Compared to SVSF, the SIF exhibited better estimation accuracy and runtime (Gadsden & Al-Shabi, 2020). Eliminating the need for analytical linearization in the SIF-based strategies for the estimation of nonlinear dynamical systems inspired by the strong tracking CKF (ST-CKF) (Zhou & Frank, 1996; Jwo & Wang, 2007), employing the cubature rule, considering fading memory concept, and reformulating the proposed gain for the SIF using the diagonal elements of the innovation covariance matrix instead of the innovation vector, the cubature SIF (CSIF) and the strong tracking innovation filter (STIF) have been introduced for the robust estimation of the nonlinear systems (Kiani & Ahmadvand, 2022). Although there are different strategies for robust estimation of nonlinear dynamical systems, all the aforementioned strategies have been presented using algorithmic approaches to implement the traditional von Neumann computer architectures. Considering the rapid evolution of various industries, there is a growing demand for estimation frameworks that simultaneously meet computational efficiency, robustness, and reliability. This need is particularly crucial in edge computing applications with resource-constrained and harsh dynamic environments such as space robotics (Flores-Abad, et al., 2014), advanced air mobility (AAM) (Johnson & Silva, 2022), and numerous other fields that necessitate size, weight, and power (SWaP) engineering (Hampo, et al., 2020). In this context, spiking neural network (SNN) architecture offers an appealing solution to address these dominant performance factors for edge computing (Schuman, et al., 2017). SNNs, characterized by their inherent scalability and minimal computational burden, present significant advantages over traditional non-spiking computational methods (Schuman, et al., 2022). SNNs represent a new generation of computing tools that leverage neural circuits composed of neurons and synapses to perform computations. In contrast to traditional non-spiking methods, which utilize sequential algorithmic approaches, SNNs draw inspiration from the brain's functioning, where neurons communicate through spikes or electrical pulses, and computation occurs when these spikes are transmitted in an asynchronous fashion (Schuman, et al., 2022; Alemi, et al., 2017; Alemi, et al., 2018; Eshraghian , et al., 2021; Yamazaki, et al., 2022). This event-driven and parallel processing nature of SNNs differs from the synchronous and sequential processing commonly found in traditional computing models. Implementing SNNs is essential for programming neuromorphic hardware, effectively serving as a programming language for edge devices (Schuman, et al., 2022). As the third generation of neural networks, SNNs employ unique neural computations that distinguish them from artificial neural networks (ANNs), reducing resource requirements in terms of computational burden and data utilization (Yamazaki, et al., 2022). In addition to computational efficiency, SNNs demonstrate remarkable scalability, making them more reliable in practical applications even under conditions akin to neuron silencing. Nevertheless, the neuron loss can be compensated by the increased spiking rate of the remaining neurons within the network (Alemi, et al., 2017). Thus, to harness the advantage of SNNs for estimation tasks, we extend a previously proposed spiking estimation strategy introduced in (Slijkhuis, et al., 2022), resembling the Luenberger observer. This approach involves the use of a constant gain that must be tailored for different systems. However, it inherits the drawbacks of the Luenberger observer, particularly its lack of robustness in the presence of uncertainties and limited ability for noise cancelation.

To address the mentioned limitations and enhance the framework introduced earlier, we first consider the recurrent network of leaky integrate and fire (LIF) neurons. Then, as a test-bed, the network weight matrices are reformulated to utilize the fully linearized EKF estimator, enabling the estimation of nonlinear systems. Additionally, using EKF empowers the proposed framework to cancel out noises effectively. To further enhance the robustness of the neuromorphic estimation framework, we extend the SNN-EKF model into another framework based on the ESIF, incorporating a gain modification technique proposed as STIF. This new framework, referred to as SNN-EMSIF, combines the computational efficiency and scalability of SNNs with the robustness of EMSIF, effectively handling bounded modeling uncertainties, which is inevitable in modeling dynamical systems. Notably, the weight matrices of the networks in SNN-EMSIF are designed according to the system model, eliminating the need for learning and estimator gain design.

Moreover, to evaluate the effectiveness of the proposed frameworks for neuromorphic filtering, a comparative analysis is conducted against the original non-spiking counterparts, utilizing a nonlinear problem called Van der Pol oscillator system with diverse types of uncertainties. Next, to assess the performance of the SNN-EMSIF in a more realistic problem, it has been applied to the state estimation of satellite rendezvous problem with partial measurements, which is a vital maneuver in space robotic applications such as on-orbit maintenance services and refueling (Flores-Abad, et al., 2014). Finally, the simulation results demonstrate the promising performance of the proposed SNN-based frameworks compared to traditional non-spiking methods in terms of accuracy, robustness, and computational efficiency.

It is noticeable that this research's primary contributions first lie in the modification of the previously observer-based spiking estimator to SNN-EKF for Kalman filtering of nonlinear dynamical systems using partially measured signals. Second, this paper proposes a new approach for implementing the EMSIF robust filtering strategy based on SNNs, which can be implemented on neuromorphic hardware, offering superior efficiency and reliability. Third, the paper investigates the proposed methods' computational burden and reliability, considering the obtained runtimes and spiking patterns. Thus, this study contributes to advancing the field of estimation theory and the advancement of neuromorphic computing methods with a focus on efficiency, robustness, and reliability.

The organization of this paper is as follows. Section 2 provides an overview of the traditional non-spiking filtering strategies, EKF and ESIF, alongside preliminaries of neuromorphic computing. We then introduce the SNN-based frameworks for Kalman filtering and robust estimation of nonlinear dynamical systems. In Section 3, numerical simulations are presented to validate the performance of the proposed methods. Finally, Section 4 concludes the paper, highlighting the contributions and potential future directions for this research in the field of estimation theory, leveraging the remarkable features of neuromorphic computing methods.

## 2. Preliminaries

This section begins by laying the groundwork and presenting some underlying preliminaries. Subsequently, the contributions of this research are outlined. In this study, the linear dynamical systems and the measurement package are assumed to adhere to the following specifications:

$$\dot{x} = f(x, u) + w \qquad (1)$$

$$z = h(x) + v \qquad (2)$$

, where $x$ is the $n_x$-dimensional state vector, $u$ is the $n_u$-dimensional control input vector, $z$ is the $n_z$-dimensional measurement vector, and $f(.)$ and $h(.)$ are nonlinear deterministic functions. $w$ and $v$ are the zero-mean Gaussian white noises with the covariance matrices of $Q$, and $R$, respectively.

### 2.1 Continuous-time extended Kalman filter (EKF)

The non-spiking time-continuous form of the EKF strategy for nonlinear deterministic systems can be considered as below (Xiao, et al., 2015):

$$\dot{\hat{x}} = f(\hat{x}, u) + K_{KF}(z - \hat{z}) \qquad (3)$$

where:

$$\hat{z} = h(\hat{x}) \quad (4)$$

$$K_{KF} = PC^T R^{-1} \quad (5)$$

$$\dot{P} = AP + PA^T + Q - PC^T R^{-1} CP \quad (6)$$

In the above equations, $A = \frac{\partial f}{\partial x}|_{x=\hat{x}}$ and $C = \frac{\partial h}{\partial x}|_{x=\hat{x}}$ are the jacobians of the dynamic process and measurement system model, respectively. The symbol ˆ denotes that the parameter has been estimated. $P$ and $K_{KF}$ refer to the state estimation error covariance matrix and Kalman filter gain, respectively.

## 2.2 Robust filtering strategy

This section briefly reviews SIF, which is the robust filtering of linear dynamical systems. The SIF framework closely resembles the KF discussed in the previous section, with the primary distinction in formulating the gain. In order to avoid redundant equations, we present only the gain formulation of the SIF below (Gadsden & Al-Shabi, 2020):

$$K_{SIF} = C^+ sat(|z - \hat{z}|/\delta) \quad (7)$$

, where $C^+$ represents the pseudo inverse of the measurement matrix, and $\delta$ refers to the sliding boundary layer, which can be tuned by trial and error. However, we adopt the gain formulation proposed in (Kiani & Ahmadvand, 2022) for the modified SIF (MSIF) filtering strategy for this research. Hence, the following expression is employed:

$$K_{MSIF} = C^+ sat(diag(P^{zz})/\delta) \quad (8)$$

, where:

$$P^{zz} = PCP^T + R \quad (9)$$

In the above expression, $P^{zz}$ refers to the innovation covariance matrix.

## 2.3 Neuromorphic computing

In order to develop a neuromorphic estimation framework, it is necessary to design a network of spiking neurons that can emulate the implementation of the estimators' dynamics. Therefore, this section will present a concise overview of the fundamental aspects of designing a network of recurrent leaky integrate-and-fire (LIF) neurons that can replicate the dynamic of fully observable linear dynamical systems. The network of LIF neurons can be defined by the following equation (Slijkhuis, et al., 2022):

$$\dot{\boldsymbol{v}} = -\lambda \boldsymbol{v} + F\boldsymbol{u}(t) + \Omega_s \boldsymbol{r} + \Omega_f \boldsymbol{s} + \boldsymbol{\eta} \quad (10)$$

where, $\boldsymbol{v} \in R^N$ is the vector of neurons membrane potential, $\lambda$ is a decay term, and $F \in R^{N \times n_u}$ matrix, which encodes the input vector. $\Omega_s \in R^{N \times N}$, and $\Omega_f \in R^{N \times N}$ are the slow and fast synaptic connections, $\boldsymbol{\eta}$ is the noise considered on the network, $\boldsymbol{r} \in R^N$ refers to the filtered spike trains, which have slower dynamics in comparison to the $\boldsymbol{s} \in R^N$ that is the emitted spike train of the neurons during each step. The dynamics of the filtered spike trains are provided below.

$$\dot{\boldsymbol{r}} = -\lambda \boldsymbol{r} + \boldsymbol{s} \quad (11)$$

Based on the theory of spike coding network (SCN) (Slijkhuis, et al., 2022; Boerlin, et al., 2013), a recurrent SNN of LIF neurons can be demonstrated for tracking the $n_x$-dimensional state vector $\boldsymbol{x}$ optimally. This tracking capability relies on two assumptions. First, the estimated state $\hat{\boldsymbol{x}}$ can be decoded from the neural activity using the following linear decoding rule:

$$\hat{\boldsymbol{x}} = D\boldsymbol{r} \quad (12)$$

where, $D \in R^{n_x \times N}$ is the random fixed decoding matrix that contains the neuron output weights. Second, the neurons have to spike only when their spike reduces the network coding error. To this aim, the firing rule of $||\boldsymbol{x} - D\boldsymbol{r}||_2^2 > ||\boldsymbol{x} - D\boldsymbol{r} - D_i||_2^2$ has to be satisfied for each neuron to emit a spike. $D_i$ is $i^{th}$ column of the matrix $D$, which could be considered as $i^{th}$

neuron output kernel that could reflect the change in the error due to a spike of $i^{th}$ neuron. This firing rule ensures that each neuron emits a spike only when it contributes to reducing the predicted error. Thus, by considering these assumptions, a recurrent network of LIF neurons capable of implementing a linear dynamical system expressed in Eq. (1) can be defined by the following expression (Slijkhuis, et al., 2022):

$$\dot{\boldsymbol{v}} = -\lambda \boldsymbol{v} + D^T(\dot{\boldsymbol{x}} + \lambda \boldsymbol{x}) - D^T D \boldsymbol{s} \quad (13)$$

This equation represents the networks' membrane potential dynamics that can be used to implement any linear dynamical system.

## 2.4 SNN-based Kalman filtering

The focus of this section is on the modifications of the previously introduced spike-based Luenberger observer (Slijkhuis, et al., 2022) by proposing a novel SNN-based KF that serves as a neuromorphic estimator, built upon the formulation provided in Eq. (13). To achieve this, with the cost of loss in the accuracy which is the consequence of leveraging jacobian matrices, we incorporate the completely linearized dynamics of the previously introduced KF presented in Eq. (3), into the expression given by Eq. (13), yielding the following modified expression:

$$\dot{\boldsymbol{v}} = -\lambda \boldsymbol{v} + D^T \left( \left( A\hat{\boldsymbol{x}} + B\boldsymbol{u} + K_{KF}(\boldsymbol{z} - \hat{\boldsymbol{z}}) \right) + \lambda \hat{\boldsymbol{x}} \right) - D^T D \boldsymbol{s} + \boldsymbol{\eta} \quad (14)$$

In the above expression $B = \frac{\partial f}{\partial \boldsymbol{u}}|_{\boldsymbol{u}=\boldsymbol{u}}$ is the Jacobian of the nonlinear dynamics concerning the input vector. Further simplification leads to the final form resembling Equation (10) with additional terms:

$$\dot{\boldsymbol{v}} = -\lambda \boldsymbol{v} + F\boldsymbol{u}(t) + \Omega_s \boldsymbol{r} + \Omega_f \boldsymbol{s} + \Omega_k \boldsymbol{r} + F_k \boldsymbol{z} + \boldsymbol{\eta} \quad (15)$$

where:

$$F = D^T B \quad (16)$$

$$\Omega_s = D^T(A + \lambda I)D \quad (17)$$

$$\Omega_f = -D^T D \quad (18)$$

Here, $\lambda$ represents the leak rate for the membrane potential, and $F$ encodes the control input to a set of spikes that is readable for the network. $\Omega_s$ and $\Omega_f$ are synaptic weights for slow and fast connections, respectively. While slow connections typically govern the implementation of the desired fully linearized estimator dynamics, in this context, they are chiefly responsible for executing the linearized dynamics of the EKF estimator, which has been presented in Eq. (3). Conversely, fast connections play a pivotal role in achieving an even distribution of the spikes across the network. Consequently, the primary contributors to the *a-priori* prediction phase of the estimation process are the second three terms in Eq. (3). In contrast, the subsequent two terms, which are influenced by $\Omega_k$, and $F_k$, adapt dynamically during the estimation process, and are tasked with handling the measurement-update or *a-posteriori* phase of the estimation. Here, $\Omega_k$ imparts the dynamics of the update component while $F_k$ furnishes the SNN with an encoded measurement vector. To update these weight matrices, the following expressions need to be used:

$$\Omega_k = -D^T(PC^T R^{-1})CD \quad (19)$$

$$F_k = D^T(PC^T R^{-1}) \quad (20)$$

The final term represents the introduced noise in the network, which imparts stochasticity to the neural activity, akin to the characteristics observed in biological neural circuits. Finally, considering the assumptions introduced in the previous section, in the proposed network, the neurons emit spikes when their membrane potentials reach their respective thresholds, defined as $T_i = (D_i^T D_i)/2$. By adhering to this firing criterion, the network incorporates the spiking behavior of neurons in the estimation process. To extract the estimated states of the considered dynamical system from the neural activity, Eq. (12) can be employed. This equation provides the estimated states from the spiking neural activity within the network. Figure 1, presents the schematic form of the non-spiking fully linearized EKF and its neuromorphic counterpart SNN-EKF. Figure 1(a) illustrates the sequential computation approach. Figure 1(b) presents the SNN-based corresponding schematic for SNN-EKF, showcasing how the sequential computation in non-spiking traditional computation methods has changed into the parallelized computation in the presented SNN, leading to the computational efficiency.

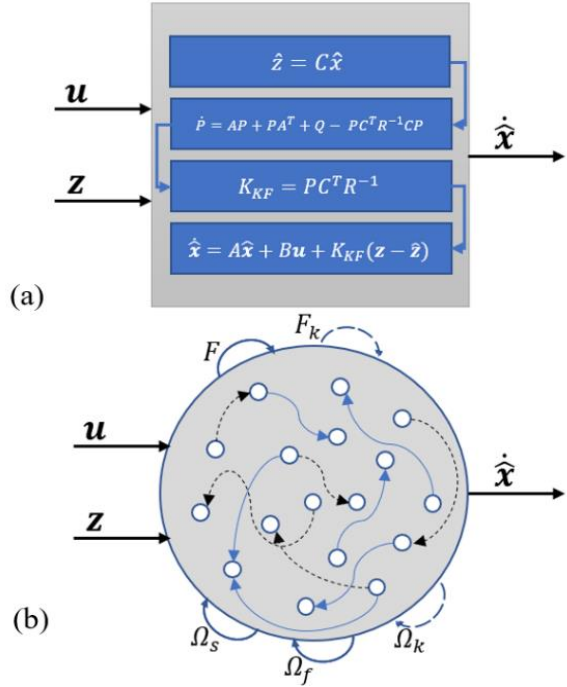

Figure 1. Schematic diagram for non-spiking EKF and SNN-based EKF, (a) non-spiking EKF, (b) SNN-based EKF

This section of the paper demonstrates the significant impact of the presented method. By adopting these approaches, we can seamlessly translate our conventional frameworks into SNNs, marking a significant milestone in favor of embracing neuromorphic computing methods over traditional sequential methods in signal processing and estimation theory. This translation facilitates the execution of equivalent tasks on neuromorphic platforms, resulting in enhanced reliability and orders of magnitude lower computational burden and energy consumption compared to conventional computers. The successful integration of traditional frameworks into SNNs paves the way for leveraging the unique characteristics of neuromorphic computing platforms, enabling substantial energy savings and improved overall performance.

2.1 **SNN-based EMSIF filtering**

In the preceding section, the SNN-based framework for the estimation of nonlinear dynamical systems, built upon the principles of the EKF has been proposed. This section aims to extend this framework to a robust filtering approach capable of handling nonlinear systems affected by bounded modeling uncertainties. To achieve this, we will reformulate Eq. (19), and Eq. (20) using the introduced gain formulation of the MSIF as presented in Eq. (8).

$$\Omega_k = -D^T(C^+ sat(diag(P^{zz})/\delta))CD \qquad (21)$$

$$F_k = D^T(C^+ sat(diag(P^{zz})/\delta)) \qquad (22)$$

Presenting the robust framework version is essential as it equips the SNN-based system to handle and mitigate uncertainties, broadening its applicability to a broader range of real-world scenarios. The gain formulation considered in this study comprises the measurement system Jacobian matrix, and the diagonal elements of the innovation covariance matrix. This formulation incorporates second-order information regarding the variations in the innovation vector, rendering it faster and more sensitive to the changes in the innovation vector (Kiani & Ahmadvand, 2022). As a result, this formulation exhibits enhanced robustness compared to its original version introduced in Eq. (7). Furthermore, utilizing this gain formulation ensures estimation stability in the presence of uncertainties (Kiani & Ahmadvand, 2022). To update the innovation covariance matrix in each time step of integration, Eq. (9), and Eq. (6) can be implemented. However, to eliminate the need for considering covariance calculation in every time step and to reduce the computational burden of the framework, the following formulations for $\Omega_k$, and $F_k$ can be employed.

$$\Omega_k = -D^T(C^+ sat(|\mathbf{z} - \hat{\mathbf{z}}|/\delta))CD \qquad (23)$$

$$F_k = D^T(C^+ sat(|\mathbf{z} - \hat{\mathbf{z}}|/\delta)) \qquad (24)$$

The above expressions are derived from the original gain formulation of SIF presented in Eq. (7). By utilizing these equations, we can effectively update the relevant weight matrices in the network and enhance its performance in the presence of uncertainties. Incorporating these updated equations represents a valuable contribution to the SNN-based estimation framework. The derived expressions enable a more efficient estimation framework with less computational burden while maintaining the system's robustness and stability. However, as discovered by (Kiani & Ahmadvand, 2022), it is worth noting that there may be a slight degradation in accuracy and sensitivity to abrupt changes in the system's states if Eq. (23) and Eq. (24) are employed. Utilizing the modified gain formulation and the derived equations contributes to the overall advancement of the SNN-based framework for the Kalman-based method, and robust estimation of nonlinear dynamical systems, especially when computational efficiency is a dominating factor in implementing such frameworks. All the simulations in this paper have been done based on Eq. (21), and Eq. (22).

## 3. Numerical simulations

In order to evaluate the performance of the proposed SNN-based frameworks in comparison to their non-spiking counterparts EKF and EMSIF, firstly, through 100 Monte-Carlo simulations, they have been applied to the estimation of a popular and highly nonlinear system called Van der Pol oscillator considering error in the initial condition, modeling uncertainty, and neuron silencing. This analysis allowed for a direct comparison of the performance between the SNN-based frameworks and traditional non-spiking approaches in a controlled setting to evaluate the applicability of the proposed methods. Subsequently, the SNN-based frameworks are applied to a more realistic problem called satellite rendezvous maneuver, considering uncertainties in systems' modeling, which is inevitable in practice. The performance, reliability, and adaptability of the SNN-based frameworks were thoroughly assessed by investigating the results.

### 3.1 Van der Pol oscillator problem

This section provides the results for the state estimation of the Van der Pol oscillator system defined by the following equations:

$$\ddot{x} - \mu(1 - x^2)\dot{x} + x = 0 \quad (25)$$

By considering the $x = [x_1, x_2] = [x, \dot{x}]$ as the state vector, the state space model $\dot{x} = f(x)$ of the differential equation presented in Eq. (25) will be drived as it has been demonstrated by the following expression.

$$\begin{bmatrix} \dot{x}_1 \\ \dot{x}_2 \end{bmatrix} = \begin{bmatrix} x_2 \\ \mu(1 - x_1^2)x_2 - x_1 \end{bmatrix} \quad (26)$$

In order to apply the EKF and EMSIF filtering strategies, the linearized model of the Van der Pol oscillator system needs to be initially driven. First, the Jacobian matrix of the presented dynamic in Eq. (26) needs to be calculated. Thus, we have the following equation system for our process and measurement system.

$$\dot{x} = \begin{bmatrix} 0 & 1 \\ -(2\mu x_1 x_2 + 1) & \mu(1 - x_1^2) \end{bmatrix} x + w \quad (27)$$

$$z = [1, 0]^T x + v \quad (28)$$

Simulations have been conducted to evaluate the performance of the proposed framework. The simulations were performed using the parameters provided in TABLE 1, with a total duration of 20 seconds and a time step of 0.01. Initially, the system is simulated without any uncertainties, followed by simulations with the inclusion of uncertainties. It is important to note that during the simulations in this section, the decoding matrix $D$ is defined using random samples from a zero-mean Gaussian distribution with a variance of 0.25. This ensured variability in the decoding matrix, reflecting real-world scenarios where the exact decoding matrix might not be known precisely and may exhibit certain uncertainties to be more realistic. The simulations assessed the robustness and performance of the proposed framework under different conditions, providing valuable insights into its effectiveness in both deterministic and uncertain scenarios. By systematically examining the impact of uncertainties on the estimation process, we comprehensively understand the framework's capabilities and limitations.

Table 1. Linear system simulation parameters

| Parameter | Value |
|---|---|
| $x_0$ | [2,2] |
| $\hat{x}_0$ | [0,0] |
| $Q$ | $I/100$ |
| $R$ | $1/10$ |
| $P_0$ | $diag([0.01, 0.01])$ |
| $N$ | 100 |
| $\lambda$ | 0.5 |
| $\delta$ | 0.05 |
| $\mu$ | 0.005 |

Figure 2 shows the results obtained from the applications of different filtering strategies. In Figure 2(a), the performance of the proposed SNN-based filters in tracking the state $x_1$ is compared. It can be observed that in contrast to the KF-based strategies, which have been converged to the true states lately almost after $t = 12$s, the EMSIF and SNN-EMSIF could track the true states in the initial time steps approximately after $t = 2$s. On the other hand, the SNN-based filters have effectively tracked the results of their non-spiking counterparts, EKF and EMSIF. Figure 2(b) compares the tracking performance of the filters for the state $x_2$. Once again, the MESIF and SNN-MESIF strategies have accurately tracked the true state with a fast convergence in almost $t = 3$s while the results obtained from EKF and SNN-EKF have converged to the true states after almost $t = 14$s. Moreover, the results demonstrated the superior performance of SNN-based strategies in tracking the results obtained by their non-spiking counterparts, which shows the validity of the proposed strategy in implementing the filters considered in this research. Here, it can be concluded that, in the presence of initial condition error, owing to their inherent robustness, the EMSIF and SNN-EMSIF outperform the EKF and SNN-EKF. Figure 2(c) and (d) depict the time histories of estimation error $x_i - \hat{x}_i$ within the $\pm 3\sigma$ bounds, which is a criterion for assessing estimation stability. Here, $\sigma_i$ represents the $i$th diagonal element of the covariance matrix $P$. In Figure 2(c), the estimation error for state $x_1$ is plotted within its respective bound. It can be observed that, because of the considered initial condition error, obtained errors for all the filters start from the outside of the bounds. The implemented filters try to reduce their estimation error to converge to a value inside the bound. In contrast to the errors obtained from EMSIF and SNN-EMSIF, which have drastically converged to the zero and entered the bound in $t = 2$s followed by a zero crossing in $t = 2.5$s with an overshot by the amount of approximately 0.2, the errors of estimates states using KF-based methods have entered to the bound at the same time as with the MSIF-based methods followed by an overshot by the amount of almost 1.8 which is nine times greater than the overshot of MSIF-based methods. Since then, the obtained errors from the EMSIF and SNN-EMSIF have approximately converged to zero in $t = 5$s while the errors obtained from the KF-based methods keep fluctuating outside the bound until $t = 10$s then they have converged to almost zero in the time period after $t = 15$s. Figure 2(d) illustrates the obtained results for the estimation errors of state $x_2$ in its respective bounds. Results demonstrate almost the same performance for the different filters in the estimation of state $x_2$. Observations here confirm that all the EMSIF-based filtering strategies employed in this simulation successfully maintained the estimation stability for the Van der Pol oscillator system, which demonstrates their robustness in keeping the estimation stability in the presence of error in the initial condition. Meanwhile, the KF-based methods exhibited a poor tracking performance.

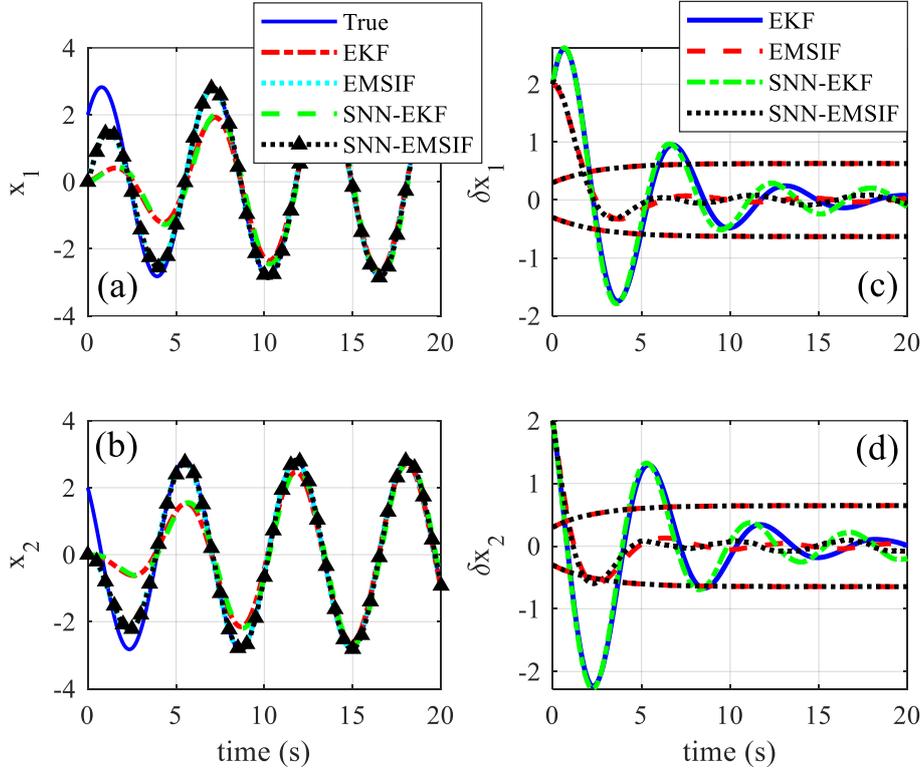

Figure 2. True and estimated states obtained from different filters along with estimation errors inside $\pm 3\sigma$ bounds, (a) estimation for the state $x_1$, (b) estimation for the state $x_2$, (c) obtained errors for state $x_1$, (d) obtained errors for state $x_2$

Figure 3 compares the time histories of root mean squared error (RMSE) obtained from different filtering methods for 100 Monte-Carlo simulations. In Figure 3(a), the results for state $x_1$ are presented, revealing the superior performance of the EMSIF in comparison to EKF in terms of estimation accuracy throughout the entire time history. Moreover, the depicted results illustrate the better performance of the SNN-EMSIF compared to the SNN-EKF in terms of estimation accuracy, demonstrating that the MSIF-based strategies are more robust in dealing with initial condition errors. Figure 3(b) shows almost the same results obtained for the estimation of the state $x_2$. On the other hand, there is another observation here, which is a loss in the accuracy of the estimation strategies when they have been translated to the SNN-based frameworks compared to their non-spiking counterparts that is the consequence of leveraging the fully linearized form implementing Jacobian of the nonlinear dynamical system. Importantly, the presented comparison also demonstrates that the SNN-based methods and traditional KF and MSIF approaches operate within the same order of magnitude in terms of estimation accuracy.

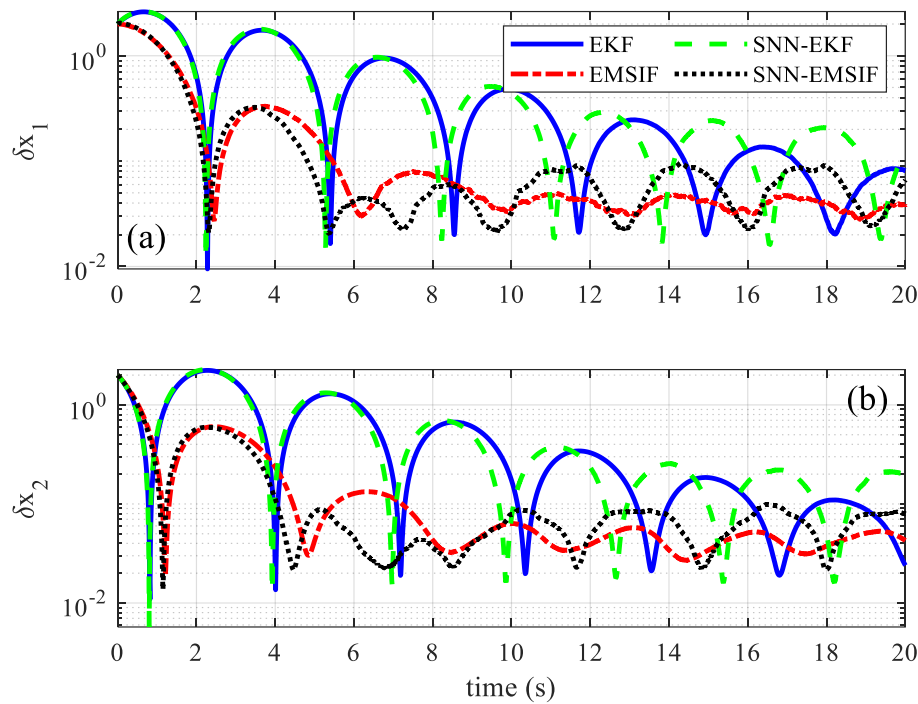

Figure 3. RMSE time-histories obtained from different filters, (a) for the state $x_1$, (b) for the state $x_2$

Moreover, to conduct a quantitative assessment of the estimation accuracies achieved by different filters, the average errors for the last 10 seconds of the simulations in which the states are approximately converged to the true values are summarized in Table 2. Remarkably, the value of reported errors for the SNN-EMSIF and EMSIF are an order of magnitude smaller than that for SNN-EKF and EKF. These results further validate the earlier observations regarding the acceptable accuracy attained by the proposed methods.

Table 2. RMSE for different filters for Van der Pol oscillator

| State | EKF | EMSIF | SNN-EKF | SNN-EMSIF |
|-------|--------|--------|---------|-----------|
| $x_1$ | 0.0166 | 0.0016 | 0.0310  | 0.0043    |
| $x_2$ | 0.0144 | 0.0020 | 0.0338  | 0.0046    |

In conclusion to this part, it can be concluded that the translated estimation strategies into neuromorphic frameworks, such as SNN-EKF for the estimation of deterministic systems and SNN-EMSIF for the systems subjected to uncertainties, can be considered as validated methods with acceptable performances for implementation on the neuromorphic hardware in neuromorphic engineering to benefit from the potential advantages of SNNs such as efficiency and scalability.

### 3.1.1 Sensitivity to Modeling Uncertainties

To assess the robustness of the proposed method in the presence of modeling uncertainties, two different cases are considered. First, by the analogy of the situations in which the process noise will be considered less than its actual value 90% of uncertainty has been considered the process noise covariance matrix implemented in the filtering frameworks by assuming it as $0.1Q$. Subsequently, with the analogy to the situation in which the measurement noise will be modeled greater than its actual value, the covariance matrix of the measurement noise is assumed to be $10R$. Then, the Monte-Carlo simulations were repeated under these new conditions. Figure 4 shows the results obtained from different filters for tracking the true state in the presence of uncertainty in the process noise modeling. In Figure 3(a), the results for the state $x_1$ are presented, demonstrating effective tracking by the filters EMSIF and SNN-EMSIF, while the filters EKF and SNN-EKF have completely deviated and could not track true values for the state $x_1$ till the end of the simulation. Figure 3(b) shows the same behavior as the state $x_1$ for the filters in the tracking of the state $x_2$ which demonstrates the KF-based filters' failure to track the true values while the results from the EMSIF and SNN-EMSIF converged to the true values in the initial time steps. These observations validate the robustness of the proposed method in the presence of uncertainty. Figures 3(c) and 3(d) illustrate the obtained estimation error time histories inside their respective $\pm 3\sigma$ bounds for states $x_1$ and $x_2$ respectively. Observations reveal that in contrast to KF-based strategies, their respective errors are highly deviated from zero and did not converge to zero within the bounds, the EMSIF and SNN-EMSIF, by keeping their respective estimation error within the bound, have exhibited a stable estimation for the states in the presence of the aforementioned uncertainty. Thus, it can be concluded that the presented framework can be satisfactorily robust and perform an estimation with acceptable accuracy in such situations.

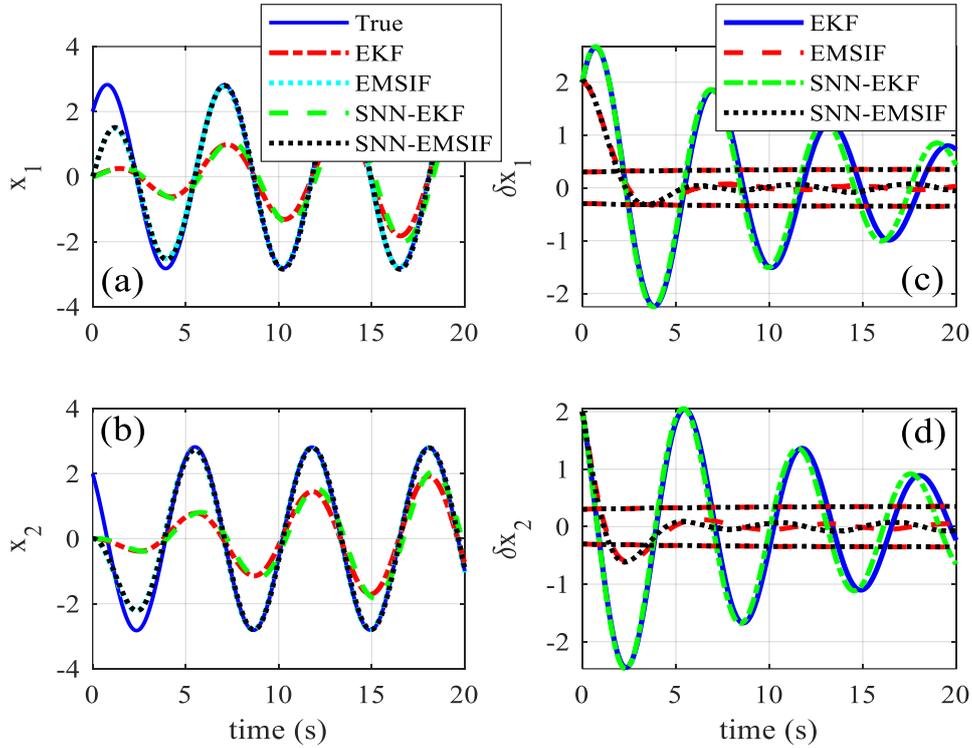

Figure 3. True and estimated states obtained from different filters along with estimation errors inside $\pm 3\sigma$ bounds for the uncertain process noise, (a) estimation for the state $x_1$, (b) estimation for the state $x_2$, (c) obtained errors for state $x_1$, (d) obtained errors for state $x_2$

Figure 4 illustrates the temporal variation of RMSEs obtained by filtering strategies in the presence of uncertainty in the process noise. In Figure 4(a), compared to the SNN-EKF and EKF, the superior Accuracy of SNN-EMSIF and its non-spiking counterpart EMSIF is observed. Conversely, the KF-based filters exhibit a rise in the obtained errors for the estimation of state $x_1$ throughout the simulation time compared to the deterministic case. Figure 4(b) shows the error variations for the state $x_2$. Again, it is demonstrated that the MSIF and SNN-MSIF consistently achieve more accurate estimations than the other methods across the time history. These results further confirm the robustness of the proposed method in the presence of uncertainty in the process noise.

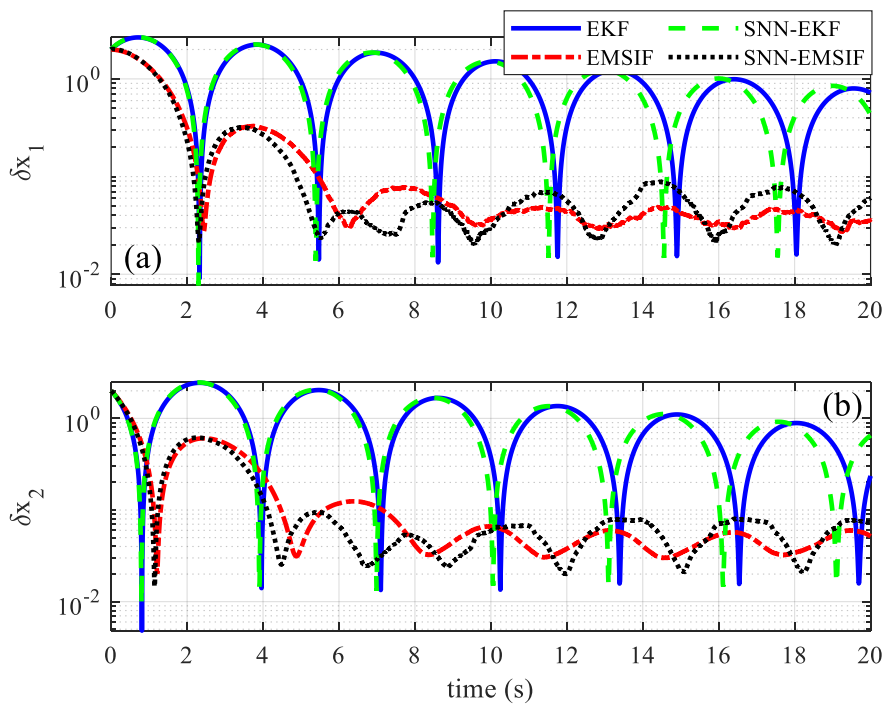

Figure 4. RMSE time-histories obtained from different filters for uncertain process noise, (a) for the state $x_1$, (b) for the state $x_2$

Furthermore, quantitatively, the average RMSEs obtained from different filters are compared in Table 3. Our results demonstrated that in the presence of the aforementioned uncertainty, the errors for KF-based methods are two orders of magnitude greater than that for SNN-EMSIF and EMSIF, which reconfirms the previous observations and demonstrates the better performance of MSIF-based methods in terms of robustness in addition to the acceptable performance of SNN-EMSIF compared to its non-spiking counterpart EMSIF.

Table 3. RMSE for different filters for Van der Pol oscillator –

Uncertain process noise

| State | EKF | EMSIF | SNN-EKF | SNN-EMSIF |
|-------|--------|--------|---------|-----------|
| $x_1$ | 0.6785 | 0.0016 | 0.6511  | 0.0031    |
| $x_2$ | 0.6093 | 0.0022 | 0.6100  | 0.0036    |

In further studies, the case of uncertain measurement noise has been considered. In this case study, because of the high deviation of KF-based filters in tracking the true values of the states and the similarity of the plots, and to avoid repetitive figures, we just provided the RMSE analysis for this case. Figure 5(a) and (b) again demonstrate the superior performance of the SNN-EMSIF and EMSIF compared to the KF-based methods in the presence of the aforementioned uncertainty for the states, $x_1$ and $x_2$, respectively. Moreover, the figure demonstrates the acceptable performance of the SNN-EMSIF compared to its non-spiking counterpart, EMSIF, in terms of accuracy and robustness in the presence of bounded modeling uncertainty. Table 4 provides the average RMSEs obtained from the filtering strategies in this study. Reported values in the table confirm our analysis of the RMSEs and again demonstrate the better performance of the SNN-EMSIF and EMSIF compared with KF-based strategies in which the errors reported for the KF-based methods are three orders of magnitude greater than that for MSIF-based frameworks. In addition, they confirm the acceptable performance of the SNN-EMSIF compared to its non-spiking counterpart, EMSIF, in terms of robustness and accuracy in the presence of the uncertainty considered in this case study.

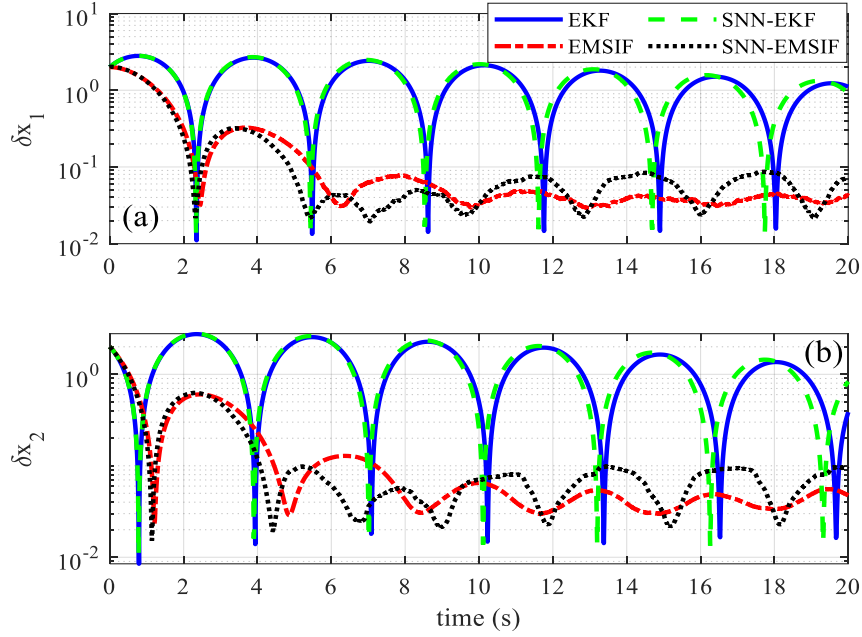

Figure 5. RMSE time-histories obtained from different filters for uncertain measurement noise, (a) for the state $x_1$, (b) for the state $x_2$

Table 4. RMSE for different filters for Van der Pol oscillator – Uncertain measurement noise

| State | EKF | EMSIF | SNN-EKF | SNN-EMSIF |
|---|---|---|---|---|
| $x_1$ | 1.4599 | 0.0015 | 1.5529 | 0.0037 |
| $x_2$ | 1.3204 | 0.0019 | 1.4307 | 0.0051 |

In the end, it can be inferred that the results obtained in the presence of the aforementioned uncertainties that are inevitable in real-world problems demonstrated the superior performance of SNN-EMSIF and EMSIF compared to the KF-based methods.

### 3.1.2 Sensitivity to neuron counts and efficiency analysis

To investigate the sensitivity of the proposed SNN-based frmaeworks to the variation of the neuron counts and the size of the network, Monte-Carlo simulations have been repeated multiple times with different neuron counts ranging from $N = 50$ to $N = 500$, increasing the neuron counts by 50 each time. The results of the simulations revealed the impact of neuron count on the performance of the estimators. Figure 6 presents the variation of the average

RMSEs on the state vector versus the neuron counts for the last 10 seconds of the considered time period. The state vector average RMSEs have been calculated using the following expression:

$$\delta_x = \sqrt{\delta_{x_1}^2 + \delta_{x_2}^2} \qquad (29)$$

The presented results demonstrate that increasing the neuron counts yields a more accurate estimation performed by the filters, exhibiting a reduction in the obtained errors. It can be observed that in contrast to region 2, in which the accuracy of the filters remained almost constant by the variations of $N$, in region 1, both the filters are sensitive to the variations of the $N$ meanwhile the SNN-EKF exhibited more sensitivity to $N$ compared to the SNN-EMSIF. On the other hand, from the perspective of the inherent scalability of the SNNs, it can be inferred that in the conditions of neuron silencing (which sometimes occurs in real-world practical applications) in which the implemented network would lose some of its neurons, the proposed frameworks are robust and keep working which is the benefit of the SNNs inherent scalability.

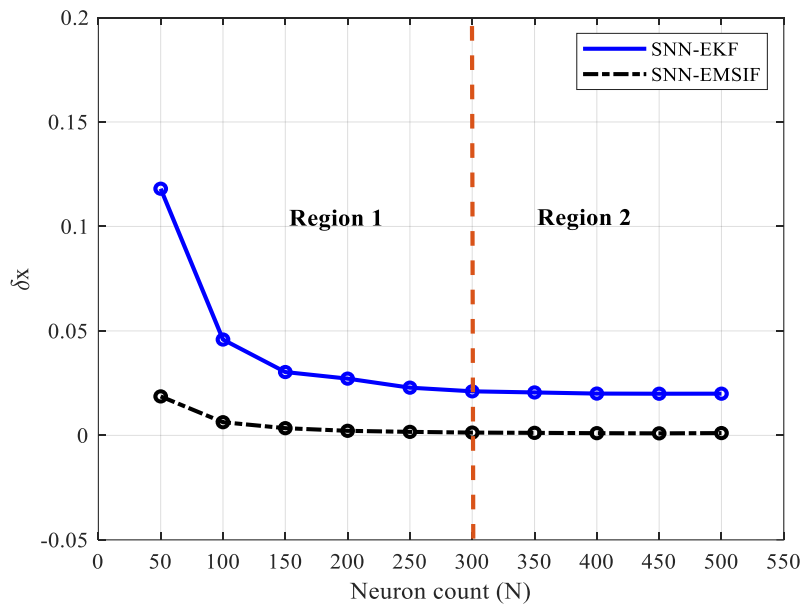

Figure 6. Obtained RMSEs versus the neuron counts for SNN-EKF and SNN-EMSIF

From an efficiency standpoint, increasing the neuron count leads to a rise in the computational burden of the designed network, making it clear that the increase in filter accuracy comes at the cost of decreased computational efficiency. Therefore, choosing the neuron count in the network design is a trade-off between accuracy, efficiency, and filter runtime. To better understand this situation, a runtime comparison between different filters was conducted. Table 5 reveals the obtained runtimes per time-step of framework implementation in MATLAB 2020b on a computer with a single Intel® Core i7-6700HQ CPU and 16 GB DDR4 memory.

Table 5. Obtained runtimes per time-step for various filters

| State | SNN-EKF | SNN-EMSIF | SNN-EMSIF$^*$ |
|---|---|---|---|
| $N = 50$ | $0.214\ ms$ | $0.219\ ms$ | $0.175\ ms$ |
| $N = 100$ | $0.300\ ms$ | $0.320\ ms$ | $0.285\ ms$ |
| $N = 300$ | $0.875\ ms$ | $0.890\ ms$ | $0.850\ ms$ |

Although the results reveal that both the employed frameworks have the runtimes of the same order, the presented runtimes demonstrated how increasing the $N$ affected the implementation runtime of the SNN-EKF and SNN-EMSIF, which is the consequence and a measure for increasing the computational burden of the network. Note that the SNN-EMSIF$^*$ in Table 5, refers to the SNN-EMSIF, which leverages Eq. (23) and Eq. (24) for updating the weight matrices of the network that perform the estimation without considering the covariance matrices. As demonstrated in Table 5, the SNN-EMSIF$^*$ has a faster implementation time. Here, it can be concluded that the choice of the neuron counts depends on the trade-off between the accuracy and the allowable computational burden, considering the limitations imposed by the problem. From the perspective of energy consumption, compared to the traditional artificial neural networks (ANN) that inherently are time-driven and all the network is active such that all the neurons are emitting signals, owing to their event-driven nature for practical implementations, the SNNs are more efficient in this way. For instance, the spiking pattern obtained from the SNN-EMSIF for a single run has been illustrated in Figure 7, in which the yellow bars represent the occurrence of spikes for neurons, and the dark background indicates the times that the membrane potential of the neurons is not charged enough to reach their thresholds, so they are not emitting signals on those times and stay standby that will cause a significant reduction in the energy consumption for practical implementation. For the spiking pattern presented in Figure 7, the SNN-EMSIF has leveraged 34,342 spikes, only 17% of all the available spikes for the 100 neurons in 2000 time steps. This means the SNN proposed here will adaptively use only 17% of the available resources in practice while the ANN will always leverage 100% of all the available resources considered for it.

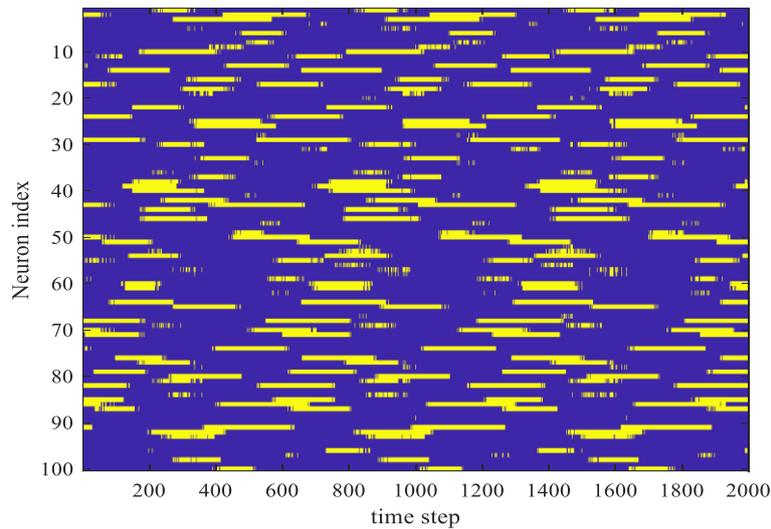

Figure 7. Obtained spiking pattern from the SNN-EMSIF for a single run

## 3.2 Satellite rendezvous problem

This section first presents the mathematical model for the satellite rendezvous maneuver, followed by the state-space model. Finally, the simulation results are provided. The rendezvous problem involves maneuvering between two distinct satellites, the chaser and the target, where the chaser approaches the target. Figure 8 schematically demonstrates this rendezvous problem.

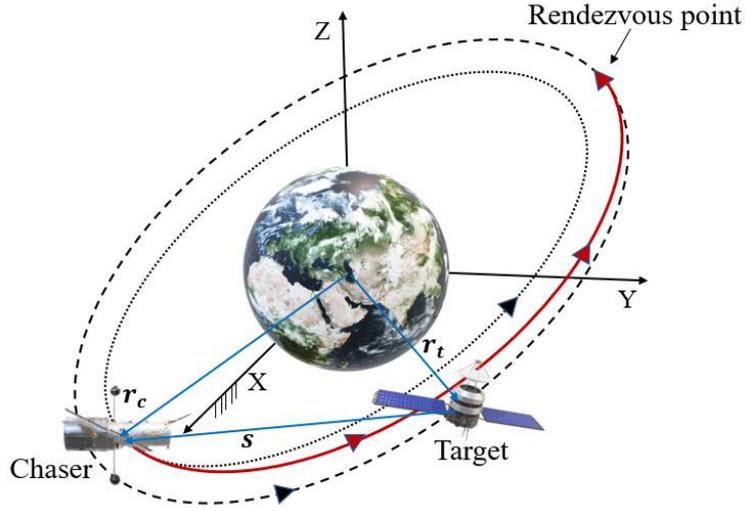

Figure 8. Schematic of rendezvous maneuver By permission from Ref. (Ahmadvand, et al., 2023)

To model the relative motion between the satellites in the Earth-centered inertial frame (ECI) the expression presented in Eq. (30) has been considered for the relative distance (Arantes & Martins-Filho, 2014).

$$\boldsymbol{s} = \boldsymbol{r}_c - \boldsymbol{r}_t \qquad (30)$$

Here, $\boldsymbol{r}_c$ and $\boldsymbol{r}_t$ are the position vectors of the chaser and target, respectively. Further, for the relative acceleration, the following equation is derived:

$$\ddot{\boldsymbol{s}} = \ddot{\boldsymbol{r}}_c - \ddot{\boldsymbol{r}}_t \qquad (31)$$

Considering the expression in Eq. (32) for the gravitational force in ECI:

$$f_g(\boldsymbol{r}) = -\mu_{earth}\frac{m}{r^3}\boldsymbol{r} \qquad (32)$$

$\mu_{earth}$ is the Earth's gravitational parameter, $m$ refers to spacecraft mass, and $\boldsymbol{r}$, and $r$ denote the spacecraft position vector and its magnitude, respectively. Thus, the absolute motion of the target and chaser satellite can separately be modeled using the following equations:

$$f_g(\boldsymbol{r}_t) = \ddot{\boldsymbol{r}}_t = -\frac{\mu_{earth}}{r_t^3}\boldsymbol{r}_t \qquad (33)$$

$$f_g(\boldsymbol{r}_c) = \ddot{\boldsymbol{r}}_c = -\frac{\mu_{earth}}{r_c^3}\boldsymbol{r}_c \qquad (34)$$

The above equations represent normalized forms of Eq. (32), divided by the spacecraft mass.

To formulate suitable equations for controller design, it is advantageous to represent relative motion in the target frame, a non-inertial reference frame rotating with the angular velocity, $\boldsymbol{\omega}$.

$$\frac{d^{*2}\boldsymbol{s}^*}{dt^2} + \boldsymbol{\omega} \times (\boldsymbol{\omega} \times \boldsymbol{s}) + 2\boldsymbol{\omega} \times \frac{d^*\boldsymbol{s}^*}{dt} + \frac{d\boldsymbol{\omega}}{dt} \times \boldsymbol{s}^* + \frac{\mu_{earth}}{r^3}M\boldsymbol{s}^* = \boldsymbol{f} \tag{35}$$

Here, $\boldsymbol{s}$ denotes relative distance, $M$, and $\boldsymbol{f}$ refer to Earth's mass and external forces, respectively, and the asterisk (*) denotes parameters in the target frame. The linearized form of Eq. (35) in the target frame, known as the Clohessy-Wiltshire (CW) equations, is expressed as (Arantes & Martins-Filho, 2014):

$$\ddot{x} - 2n\dot{z} = f_x \tag{36}$$

$$\ddot{y} + n^2\dot{y} = f_y \tag{37}$$

$$\ddot{z} + 2n\dot{x} - 2n^2\dot{z} = f_z \tag{38}$$

where:

$$n = \sqrt{\frac{\mu_{earth}}{R_o^3}} \tag{39}$$

Here, $R_o$ represents the orbital radius of the target spacecraft, and $n$ is the mean motion. To design the LQR controller, we begin by defining the state and input vectors as $\boldsymbol{x} = [x, y, z, \dot{x}, \dot{y}, \dot{z}]^T$, and $\boldsymbol{u} = [f_x, f_y, f_z]$, respectively. Subsequently, we derive the state space form of CW equations, expressed as:

$$\dot{\boldsymbol{x}} = A\boldsymbol{x} + B\boldsymbol{u} \tag{40}$$

where:

$$A = \begin{bmatrix} 0 & 0 & 0 & 1 & 0 & 0 \\ 0 & 0 & 0 & 0 & 1 & 0 \\ 0 & 0 & 0 & 0 & 0 & 1 \\ 0 & 0 & 0 & 0 & 0 & 2n \\ 0 & 0 & 0 & 0 & -n^2 & 0 \\ 0 & 0 & 0 & -2n & 0 & 2n^2 \end{bmatrix} ; B = \begin{bmatrix} 0 & 0 & 0 \\ 0 & 0 & 0 \\ 0 & 0 & 0 \\ 1 & 0 & 0 \\ 0 & 1 & 0 \\ 0 & 0 & 1 \end{bmatrix} \tag{41}$$

$$\boldsymbol{u} = -K\hat{\boldsymbol{x}} \tag{42}$$

Here, the symbol $\widehat{\phantom{x}}$, denotes an estimated parameter. And $K$ refers to controller gain, which can be designed independently, which is out of the scope of this research. The simulations in this section are conducted using the numerical values provided in Table 6, with a time duration of 360 seconds and a time step of 0.1. Additionally, the decoding matrix $D$ is defined using random samples from zero-mean Gaussian distributions with covariances of 1/15. Note that the simulations here have been conducted in various types of uncertainty. To add the uncertainty, in the leveraged model in the filters for the noise covariance matrices, the values of $\widehat{Q} = 0.9Q$ and $\widehat{R} = 5R$ have been considered in the estimation frameworks.

Table 6 Parameters for satellite rendezvous problem

| Parameter | Value |
|---|---|
| $r_0$ (m) | $[70, 30, -5]^T$ |
| $v_0$ (m/s) | $[-1.7, -0.9, 0.25]^T$ |
| $x_0$ | $[r_0, v_0]^T$ |
| $\widehat{x}_0$ | $x_0$ |
| $Q$ | $(1e-12)I_6$ |
| $R$ | $(1e-2)I_3$ |
| $N$ | 200 |
| $\lambda$ | 0.001 |
| $\mu_{earth}$ | 398600 |
| $\delta$ | 0.1 |

Figure 9 shows the estimated states obtained from different filters compared to true states. Obtained results demonstrate the better tracking performance of the SNN-EMSIF filtering strategy such that in contrast to the obtained results from SNN-EMSIF such that they have exhibited good tracking performance of the true values, the results obtained from SNN-EKF have slightly deviated from the true value. Particularly for the states $z$ and $v_z$ in which the deviations are observable.

Figure 10 compares the obtained estimation errors within $\pm 3\sigma$ bounds for SNN-EKF and SNN-EMSIF. Comparisons demonstrated that the errors from the SNN-EMSIF converged to almost zero with minor fluctuations, while the results from the SNN-EKF failed to converge to zero and continued fluctuating around zero, with significantly larger deviations compared to SNN-EMSIF. Thus, it can be inferred that in the presence of uncertainties, the SNN-EMSIF has demonstrated better accuracy, confirming the proposed method's robustness. On the other hand, it can be observed that the errors obtained in both filters remained in the bouns. It can be interpreted that both the filters performed a stable estimation but with different accuracies.

Figure 11 illustrates the RMSEs obtained from different filters, demonstrating that in the presence of modeling uncertainty, SNN-MSIF outperforms the SNN-EKF estimator in terms of estimation accuracy. Table 7 also quantitatively compares the average RMSEs obtained from different filters in the last 60 seconds of the simulation, where the errors have almost converged to zero. The results reported in the table reaffirm the findings from Figures 10 and 11.

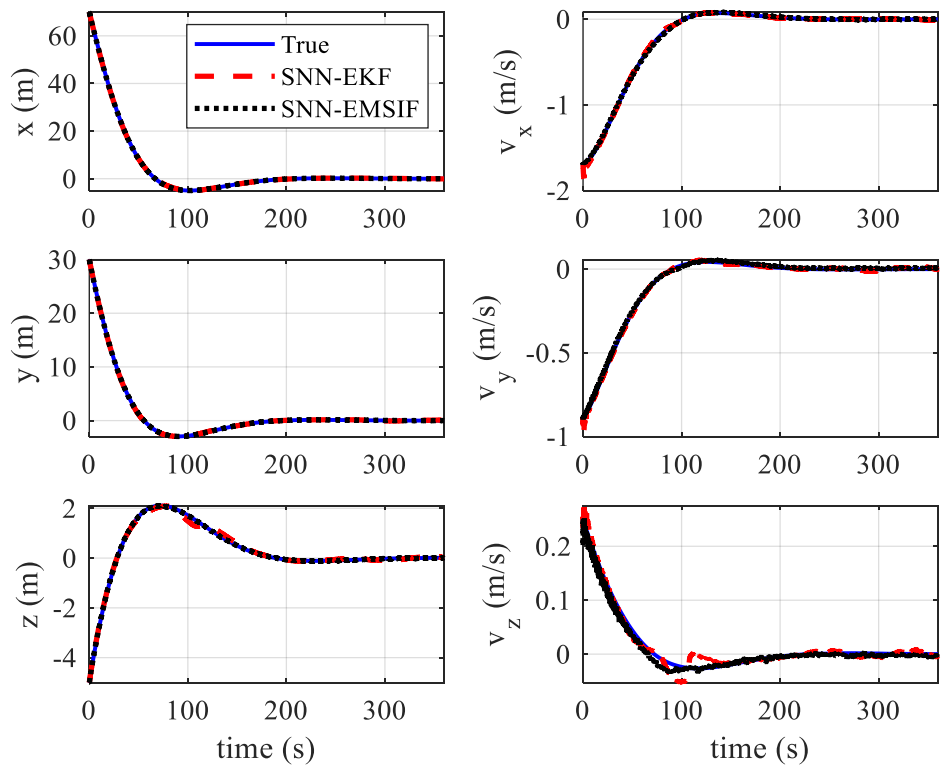

Figure 9. Time histories of the estimated states in the rendezvous problem

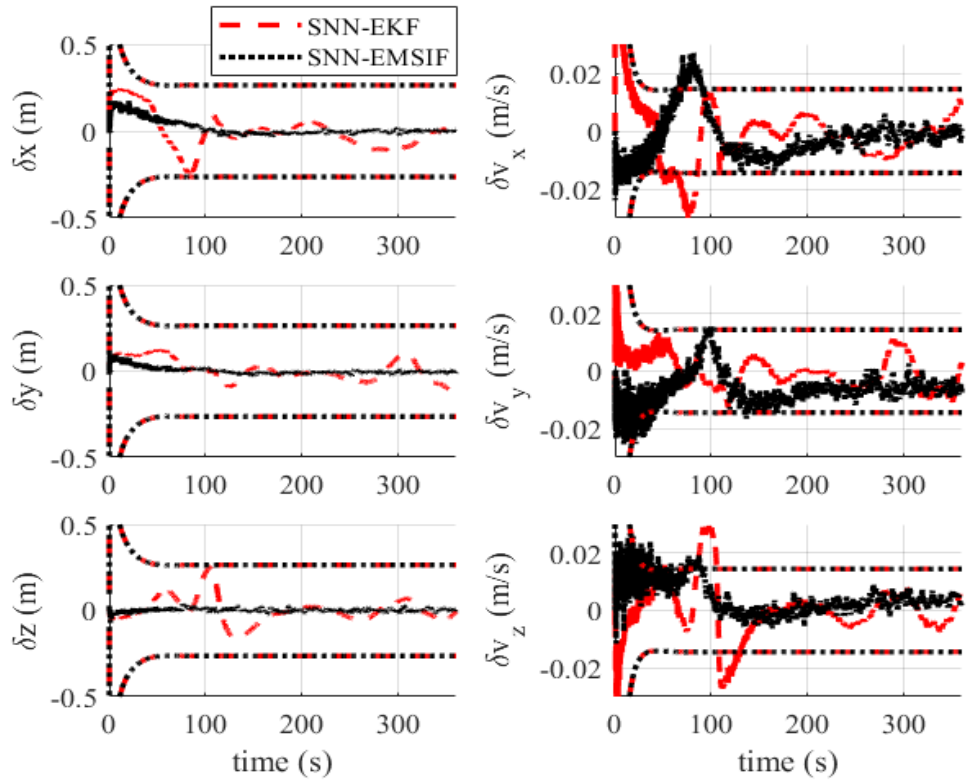

Figure 10. Obtained estimation errors for various filters within $\pm 3\sigma$ bounds in rendezvous problem.

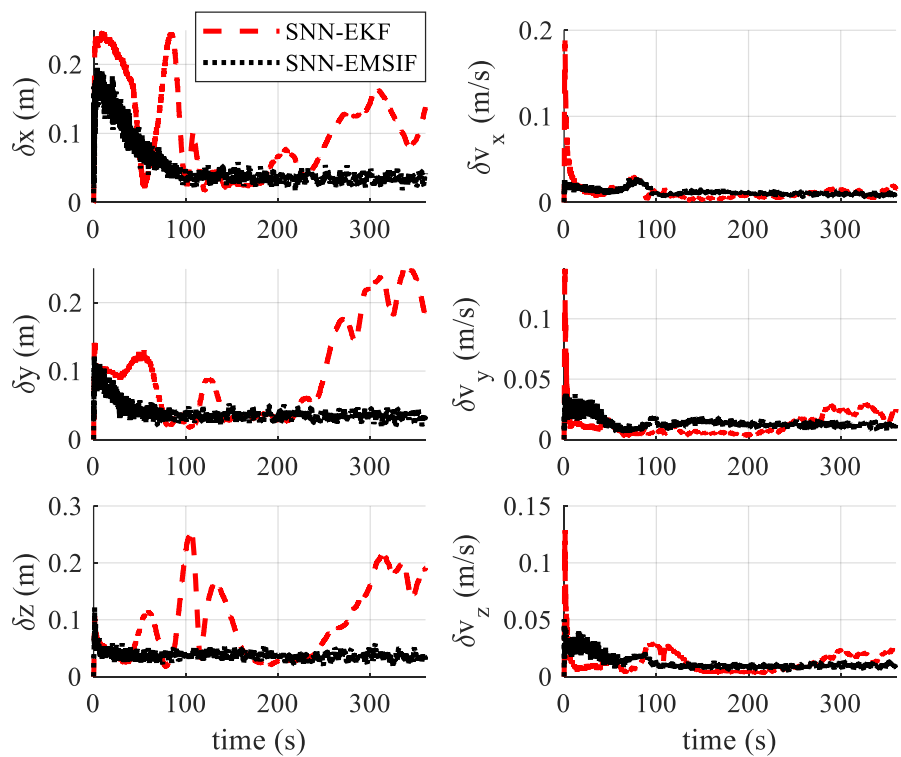

Figure 11. Obtained RMSEs for different filters in the rendezvous problem

Table 7. Average RMSE obtained from different filters
for rendezvous problem

| State | SNN-EKF | SNN-EMSIF |
|---|---|---|
| $x\ (m)$ | 0.0157 | 0.0013 |
| $y\ (m)$ | 0.0501 | 0.0012 |
| $z\ (m)$ | 0.0342 | 0.0013 |
| $v_x\ (m/s)$ | 1.5169e-04 | 9.2758e-05 |
| $v_y\ (m/s)$ | 5.4087e-04 | 1.4348e-04 |
| $v_z\ (m/s)$ | 3.9759e-04 | 9.8731e-05 |

Obtained results in this section demonstrated that the proposed estimation framework SNN-EMSIF can be considered a good and valid strategy for the implementation in real-world space robotic problems in which the modeling uncertainty is an inevitable thing and simultaneously benefits from the advantages of neuromorphic computing in the space industry that constantly demands efficient, scalable and reliable frameworks.

## 4. Conclusion

Throughout this study, we have strongly emphasized the crucial factors of computational efficiency, robustness, and reliability in the state estimation of nonlinear dynamical systems. The widespread application and significance of the filtering strategies in these domains have also been thoroughly investigated. Considering recent advances in neuromorphic computing approaches across scientific fields, this study has focused on introducing a neuromorphic approach for implementing a robust estimation framework for nonlinear dynamical systems affected by modeling uncertainties using partially measured signals. To address this, we presented the SNN-EMSIF framework, a new approach based on SNNs, and the EMSIF, which is a robust filtering strategy. The proposed framework benefits from the computational efficiency and scalability of SNNs and the EMSIF's robustness. To this aim, an SNN-EKF has been introduced for the Kalman filtering of nonlinear dynamical systems with zero-mean

Gaussian noises. Further, to demonstrate the validity of these methods, we compare them with their non-spiking counterparts, namely, EKF and EMSIF. Our comprehensive Monte-Carlo simulations illustrate the superior performance and robustness of the SNN-EMSIF when compared to the SNN-EKF, particularly in the presence of uncertainties. One of the key advantages of our proposed frameworks lies in leveraging the computational efficiency and scalability of neuromorphic computing methods. In addition, based on their event-based nature and highly parallelized computation, these frameworks offer significant advantages over traditional non-spiking sequential methods and ANNs for equivalent tasks.

In conclusion, our research showcases the immense potential of the proposed methods in robust estimation of nonlinear systems. In future work, we aim to apply the presented frameworks in this research to the sensor fusion problem, which is crucial for state estimation and localization problems that leverage various types of sensors. The promising outcomes of this study encourage further exploration and integration of neuromorphic computing in various domains, facilitating efficient and reliable solutions for complex real-world challenges.